\documentclass{llncs}

\usepackage{graphicx}
\usepackage{hyperref}
\usepackage{cite}

\begin{document}

\title{Complexity, Development, and Evolution in Morphogenetic Collective Systems}
\author{Hiroki Sayama}
\institute{Center for Collective Dynamics of Complex Systems\\
Department of Systems Science and Industrial Engineering\\
Binghamton University, State University of New York\\
\email{sayama@binghamton.edu}}

\maketitle

\begin{abstract}
Many living and non-living complex systems can be modeled and
understood as collective systems made of heterogeneous components that
self-organize and generate nontrivial morphological structures and
behaviors. This chapter presents a brief overview of our recent effort
that investigated various aspects of such morphogenetic collective
systems. We first propose a theoretical classification scheme that
distinguishes four complexity levels of morphogenetic collective
systems based on the nature of their components and interactions. We
conducted a series of computational experiments using a self-propelled
particle swarm model to investigate the effects of (1) heterogeneity
of components, (2) differentiation/re-differentiation of components,
and (3) local information sharing among components, on the
self-organization of a collective system. Results showed that (a)
heterogeneity of components had a strong impact on the system's
structure and behavior, (b) dynamic differentiation/re-differentiation
of components and local information sharing helped the system maintain
spatially adjacent, coherent organization, (c) dynamic
differentiation/re-differentiation contributed to the development of
more diverse structures and behaviors, and (d) stochastic
re-differentiation of components naturally realized a self-repair
capability of self-organizing morphologies. We also explored
evolutionary methods to design novel self-organizing patterns, using
interactive evolutionary computation and spontaneous evolution within
an artificial ecosystem. These self-organizing patterns were found to
be remarkably robust against dimensional changes from 2D to 3D,
although evolution worked efficiently only in 2D settings.
\end{abstract}

\section{Introduction}

Various living and non-living systems are collective systems in the
sense that they consist of a large number of smaller
components. Those microscopic components interact with each other to
show a wide variety of self-organizing macroscopic structures and
behaviors, which have been subject to many scientific inquiries
\cite{baryam1997,benjacob1998,parrish1999,sole2000,macy2002,camazine2003,couzin2003,gershenson2007,lammer2008,turner2008,turner2011,vicsek2012,portugali2012,doursat2012,fernandez2014,sayama2015textbook}.

Typical assumptions often made in earlier mathematical/computational
models of self-organizing collectives include the homogeneity of
individual components' properties and behavioral rules within a
collective. Such homogeneity assumptions have merit in simplifying
models and allowing for analytical prediction of the models'
macroscopic behaviors. However, such homogeneity assumptions would not
be adequate to capture more complex nature observed in real-world
complex systems, such as multi-cellular organisms' morphogenesis and
physiology \cite{sole2000,camazine2003}, termite colony building and
maintenance \cite{turner2008,turner2011}, and growth and
self-organization of human social systems
\cite{macy2002,lammer2008}. Those real-world complex collectives
consist of heterogeneous components whose behavioral types can change
dynamically via active information exchange among locally connected
neighbors. These properties of components facilitate self-organization
of highly nontrivial morphological structures and behaviors
\cite{sayama2014a}.

In this chapter, we present a brief summary of our recent effort in
investigating several aspects of complex morphogenetic collective
systems that involve (1) heterogeneous components, (2) dynamic
differentiation/re-differentiation of the components, and (3) local
information sharing among the components. Our objective was to
understand the implications of each of those properties for
developmental processes of the collectives, and to develop effective
methodologies to design novel artificial morphogenetic collective
systems.

The rest of this chapter is structured roughly following the topics of
this proceedings volume---{\em evolution, development, and
  complexity}---though we will discuss them in a reversed order. We
will first propose a classification scheme of several distinct
complexity levels of morphogenetic collective systems based on their
components' functionalities. Then we will computationally investigate
how the developmental processes, i.e., self-organization of
morphological patterns created by interacting components, will be
affected by the difference in the complexity levels of those
systems. Finally, we will discuss evolutionary methods to design
nontrivial self-organization of morphogenetic collective systems, with
a brief additional remark on their robustness/sensitivity to spatial
dimensional changes.

\section{Functional Complexity Levels of Morphogenetic Collective Systems}

Our first task is to identify what kind of properties are typically
seen in real-world complex collective systems but often omitted for
simplicity in the literature on mathematical/computational models of
those systems. In \cite{sayama2014a}, we selected the following three
as the key properties essential for self-organization of morphogenetic
collective systems yet often ignored in the literature:
\begin{enumerate}
\item Heterogeneity of components
\item Differentiation/re-differentiation of components
\item Local information sharing among components
\end{enumerate}

Heterogeneity of components means that there are multiple, distinct
types of components whose behaviors are different from each
other. Note that these types are not necessarily a simple rewording of
dynamical states. Instead, each type may have multiple dynamical
states within itself, while its behavioral rules as a whole (e.g.,
state-transition rules) should be different from those of other
types. Examples include different cell types within an organism,
individuals with different phenotypical traits in a colony of social
insects, and different professions of individuals in human
society. Differentiation/re-differentiation means that each individual
component will assume one of those types (differentiation), and
potentially switch from one type to another under certain conditions
(re-differentiation). Finally, local information sharing means that
the individual components are actively sending/receiving encoded
signals among them for coordination of their collective behaviors,
such as cell-cell communication with molecular signals,
pheromone-based communication among social insects, and human
communication in languages.

\begin{figure}[t]
\centering
\includegraphics[width=0.8\columnwidth]{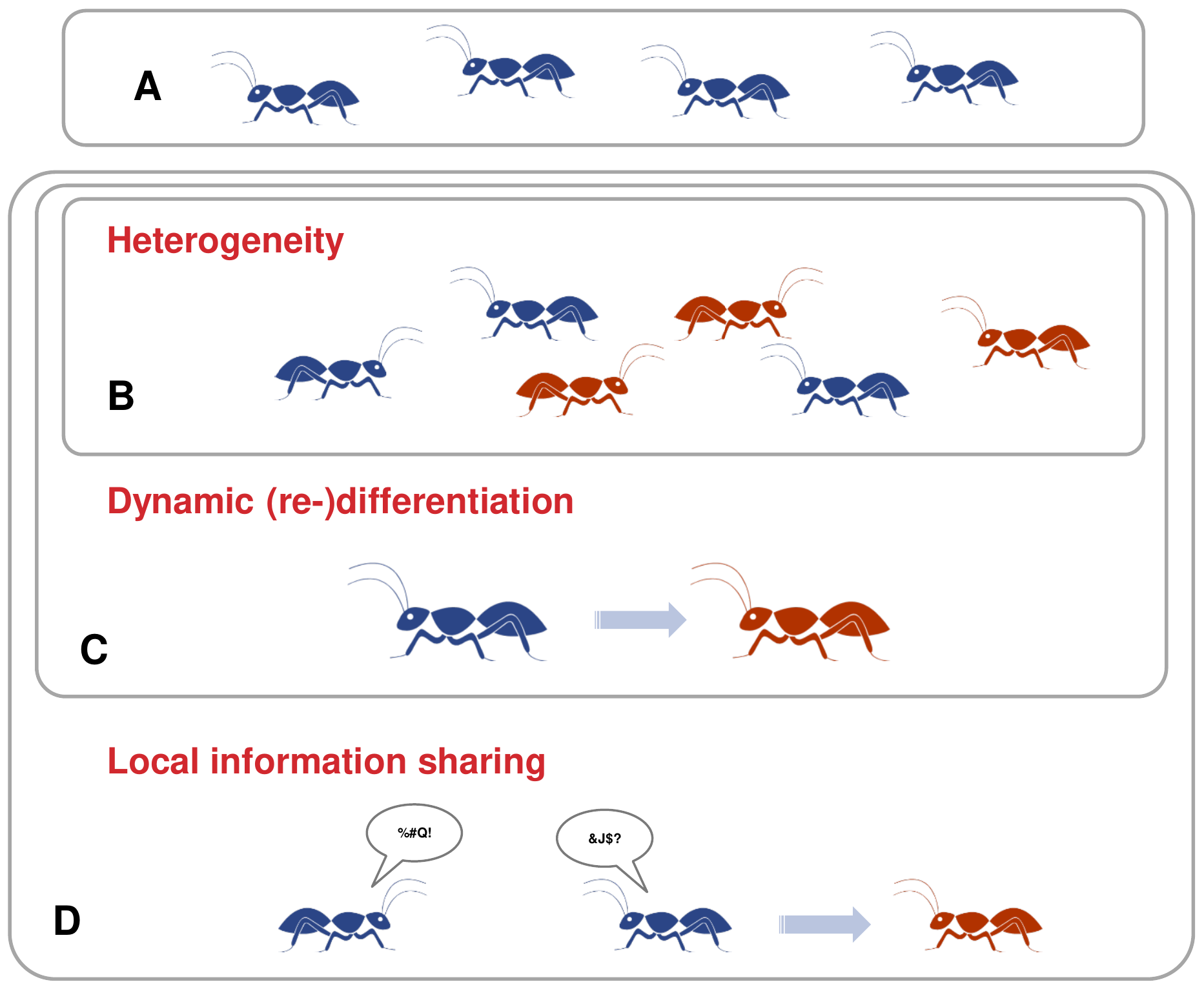}
\caption{Proposed four levels of complexity of morphogenetic
  collective systems. A: Homogeneous collective. B: Heterogeneous
  collective. C: Heterogeneous collective with dynamic
  (re-)differentiation. D: Heterogeneous collective with dynamic
  (re-)differentiation and local information sharing. These four
  classes form a hierarchical level structure; see text for details.}
\label{fig:four-classes}
\end{figure}

Mathematically speaking, distinguishing presence/absence of each of
these three properties would define a total of $2^3 = 8$ possible
classes of collective systems. However, we claim that there are some
hierarchical relationships among those three properties. Specifically,
differentiation/re-differentiation of components require, almost
tautologically, the multiple possibilities of component
types. Furthermore, we assumed that local information sharing would
make sense only if the components had an ability to change their types
dynamically based on the received information\footnote{We note that
  this assumption is much less obvious than the first one, and if we
  did not adopt it, we would obtain $3 \times 2 = 6$ different
  classes. In this chapter, we limit our focus on the four-level
  classification presented above.}. Taking these requirement
relationships into account, we proposed the following four
hierarchical classes of complexity levels of morphogenetic collective
systems \cite{sayama2014a} (Fig.~\ref{fig:four-classes}):
\begin{description}
\item [Class A] Homogeneous collective
\item [Class B] Heterogeneous collective
\item [Class C] Heterogeneous collective with dynamic (re-)differentiation
\item [Class D] Heterogeneous collective with dynamic (re-)differentiation and local
  information sharing
\end{description}
The dynamics of components in each of these four classes can be
represented mathematically as follows \cite{sayama2014a}:
\begin{description}
\item[Class A] $a^{t+1}_i = F(o^t_i)$\\
\item[Class B] $a^{t+1}_i = F(s_i , o^t_i)$\\
\item[Class C] $a^{t+1}_i = F(s^t_i , o^t_i)$, $s^{t+1}_i = G(s^t_i , o^t_i)$\\
\item[Class D] $a^{t+1}_i = F(S^t_i , O^t_i)$, $s^{t+1}_i = G(S^t_i , O^t_i)$,\\
$S^t_i = \{s^t_m \; | \; m \mathrm{~in~} N^t_i \}$, $O^t_i = \{o^t_m \; | \; m \mathrm{~in~} N^t_i \}$
\end{description}
Here $a^t_i$, $o^t_i$, and $s^t_i$ are individual component $i$'s
behavior, observation, and type at time $t$, respectively ($s_i$ is a
time-invariant type of component $i$); $F$ and $G$ are model
functions; and $N^t_i$ is the set of component $i$'s neighbors at time
$t$. These mathematical formulations help clarify the hierarchical
relationships among the four complexity levels. Following these
formulations, we will construct a specific computational model of
morphogenetic collective systems to facilitate systematic
investigation of the proposed four complexity levels and their
characteristics.

\section{Developmental Models: Morphogenetic Swarm Chemistry}

\begin{figure}[t]
\centering
\includegraphics[width=\columnwidth]{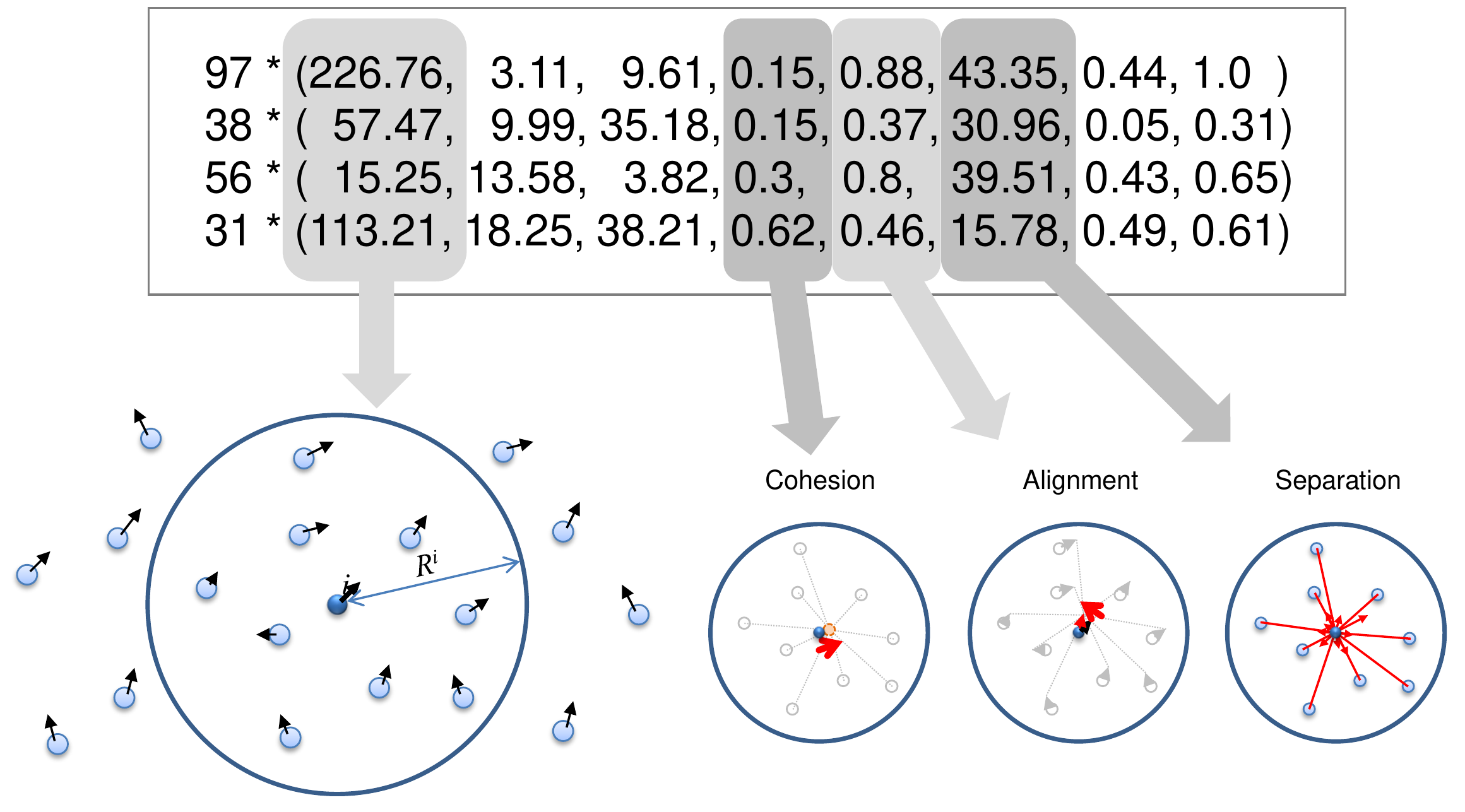}
\caption{Encoding of behavioral parameters in a {\em recipe} in Swarm
  Chemistry. A recipe is a list of parameter values written in the
  format ``{\em number of particles} * ({\em parameter values for
    behaviors of those particles})''. The parameters include the
  radius of interaction (bottom left) and the strengths of three
  primary rules (cohesion, alignment, and separation; bottom right).}
\label{fig:recipe}
\end{figure}

We utilized our earlier ``Swarm Chemistry'' model
\cite{sayama2009a,sayama2012a} to construct a new computational model
of morphogenetic collective systems. Swarm Chemistry is a revised
version of Reynolds' well-known self-propelled particle swarm model
known as ``Boids'' \cite{reynolds1987}. In Swarm Chemistry, multiple
types of components with different kinetic behavioral parameters are
mixed together. Their behavioral parameters are represented in a
``recipe'' as shown in Fig.~\ref{fig:recipe}. Therefore, the Swarm
Chemistry model is already capable of representing both Class A
(homogeneous) and Class B (heterogeneous) collective systems. In Swarm
Chemistry, components with different types spontaneously segregate
from each other even without any sophisticated sensing or control
mechanisms, often forming very intricate self-organizing dynamic
patterns \cite{sayama2009a,sayama2012a}.

\begin{figure}[t]
\centering
\includegraphics[width=\columnwidth]{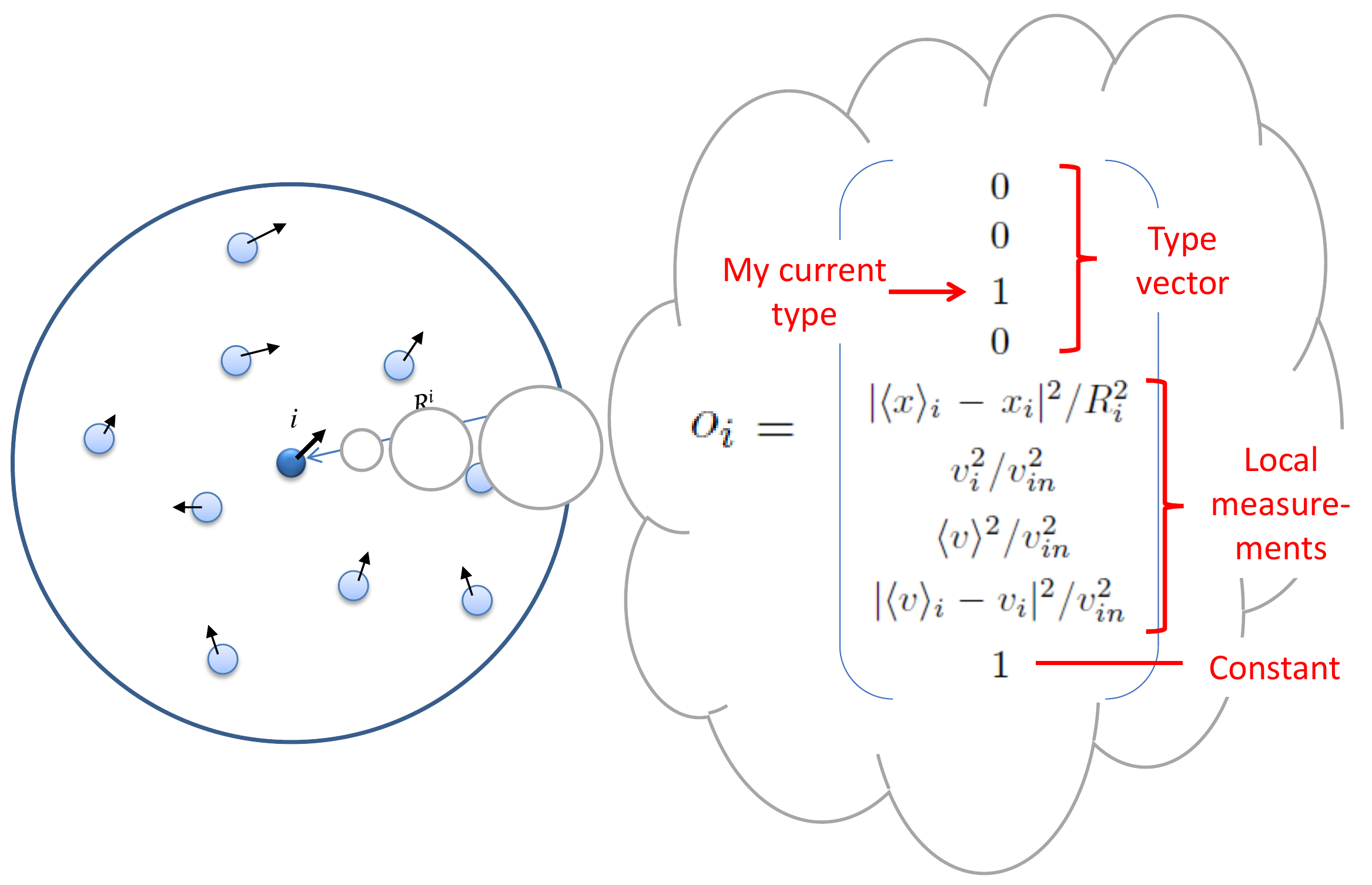}
\caption{Observation vector $o$ of each particle used in Morphogenetic
  Swarm Chemistry. The first several values of $o$ encode the current
  type of the particle, while the rest captures the measurements of
  its local environment. A constant unity is also included at the end
  of the vector.}
\label{fig:observation-vector}
\end{figure}

To make individual components capable of dynamic
differentiation/re-differentiation and local information sharing, we
made several extensions to Swarm Chemistry \cite{sayama2014a}. First,
we made each individual component able to obtain information about its
own dynamical type and its local environment in the form of {\em
  observation vector} $o$ (Fig.~\ref{fig:observation-vector}), and
then utilize this vector to decide which dynamical type it should
assume. This allows for dynamic (re-)differentiation required for
Class C/D collective systems. This decision making process was
implemented via multiplication of {\em preference weight matrix} $U$
to the observation vector $o$, so that letting $U=0$ represents Class
A/B systems as well. The second model extension was to introduce {\em
  local information sharing coefficient} $w$, with which the actual
input vector multiplied by $U$ was calculated as the weighted average
between the component's own observation vector and the local average
of all the observation vectors of neighbor components. Changing the
value of $w$ represents switching between Class C and Class D
collective systems. With these, the four complexity levels discussed
in the previous section were fully parameterized as shown in Table
\ref{tab:parameterization}. This expanded model is called
``Morphogenetic Swarm Chemistry'' hereafter. More details can be found
in \cite{sayama2014a}.

\begin{table}[t]
\centering
\caption{Parameterization of four complexity levels of Morphogenetic
  Swarm Chemistry models.}
\begin{tabular}{l|llllll}
\hline
Class & & Recipe & & $U$ & & $w$\\
\hline
A & & Single-type &  & $0$ & & $0$\\
B & & Multiple-type & & $0$ & & $0$\\
C & & Multiple-type & & $\ne 0$ & & $0$\\
D & & Multiple-type & & $\ne 0$ & & $\ne 0$\\
\hline
\end{tabular}
\label{tab:parameterization}
\end{table}

\section{Differences of Developmental Processes Across Complexity Levels}

We conducted a series of computational experiments using the
Morphogenetic Swarm Chemistry model to investigate the differences of
their developmental processes across the four complexity levels. This
was conducted by detecting statistical differences in topologies and
behaviors of self-organizing patterns that were collected via Monte
Carlo simulations using randomly sampled parameter values. Topological
and behavioral features of self-organizing patterns were measured
using several kinetic metrics (average speed, average absolute speed,
average angular velocity, average distance from center of mass,
average pairwise distance) as well as newly developed network
analysis-based metrics
\cite{sayama2015textbook,wasserman1994,barabasi2016} (number of
connected components, average size of connected components,
homogeneity of sizes of connected components, size of largest
connected component, average size of non-largest connected components,
average clustering coefficient, link density) that were measured on a
network reconstructed from the individual components' positions in
space \cite{sayama2014a}. These new metrics allowed us to capture
topological properties of the collectives that would not have been
captured by using simple kinetic metrics only.

\begin{figure}[t]
\centering
\includegraphics[width=\columnwidth]{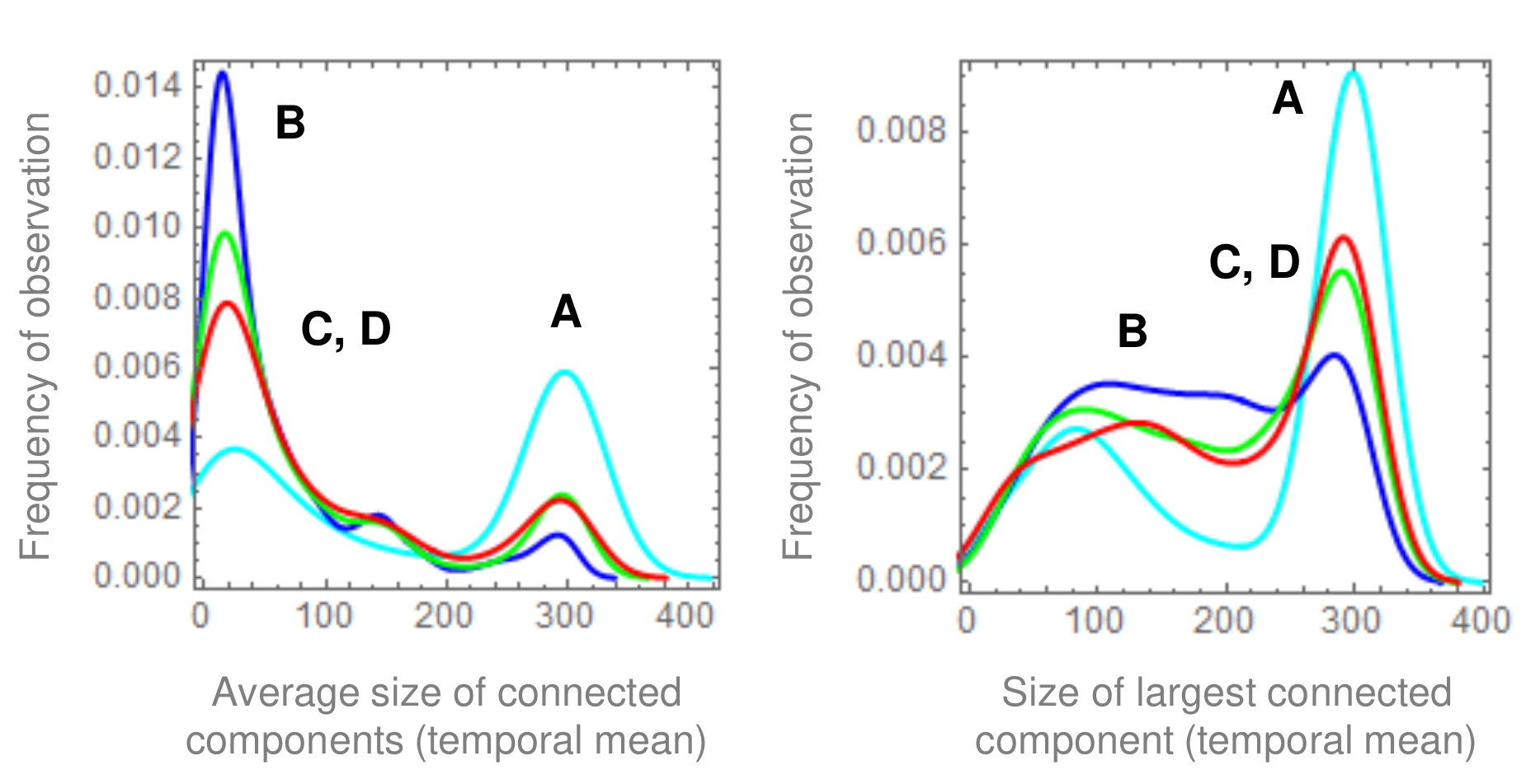}
\caption{Examples of experimental results showing clear differences of
  morphological properties among the four classes. Left: Distributions
  of the average size of connected components in generated
  morphologies. Right: Distributions of the size of the largest
  connected component in generated morphologies. In both plots,
  Classes C and D show intermediate distributions between those of
  Class A and Class B.}
\label{fig:findings}
\end{figure}

Results showed significant differences in most of the metrics between
the four different classes of morphogenetic collective systems
\cite{sayama2014a}. Specifically, heterogeneity of components had a
strong impact on the system's structure and behavior, and dynamic
differentiation/re-differentiation of components and local information
sharing helped the system maintain spatially adjacent, coherent
organization. Statistical differences were particularly significant
for topological features, demonstrating the effectiveness of our newly
developed network analysis-based metrics. It was also observed that
the properties of Class C/D collective systems tended to fall in
between Class A and Class B in many metrics
(Fig.~\ref{fig:findings}). Moreover, it was noted that, as a
byproduct, stochastic re-differentiation of components naturally
realized a self-repair capability of self-organizing morphologies
\cite{sayama2012a,sayama2010}.

\begin{figure}[t]
\centering
\includegraphics[width=\columnwidth]{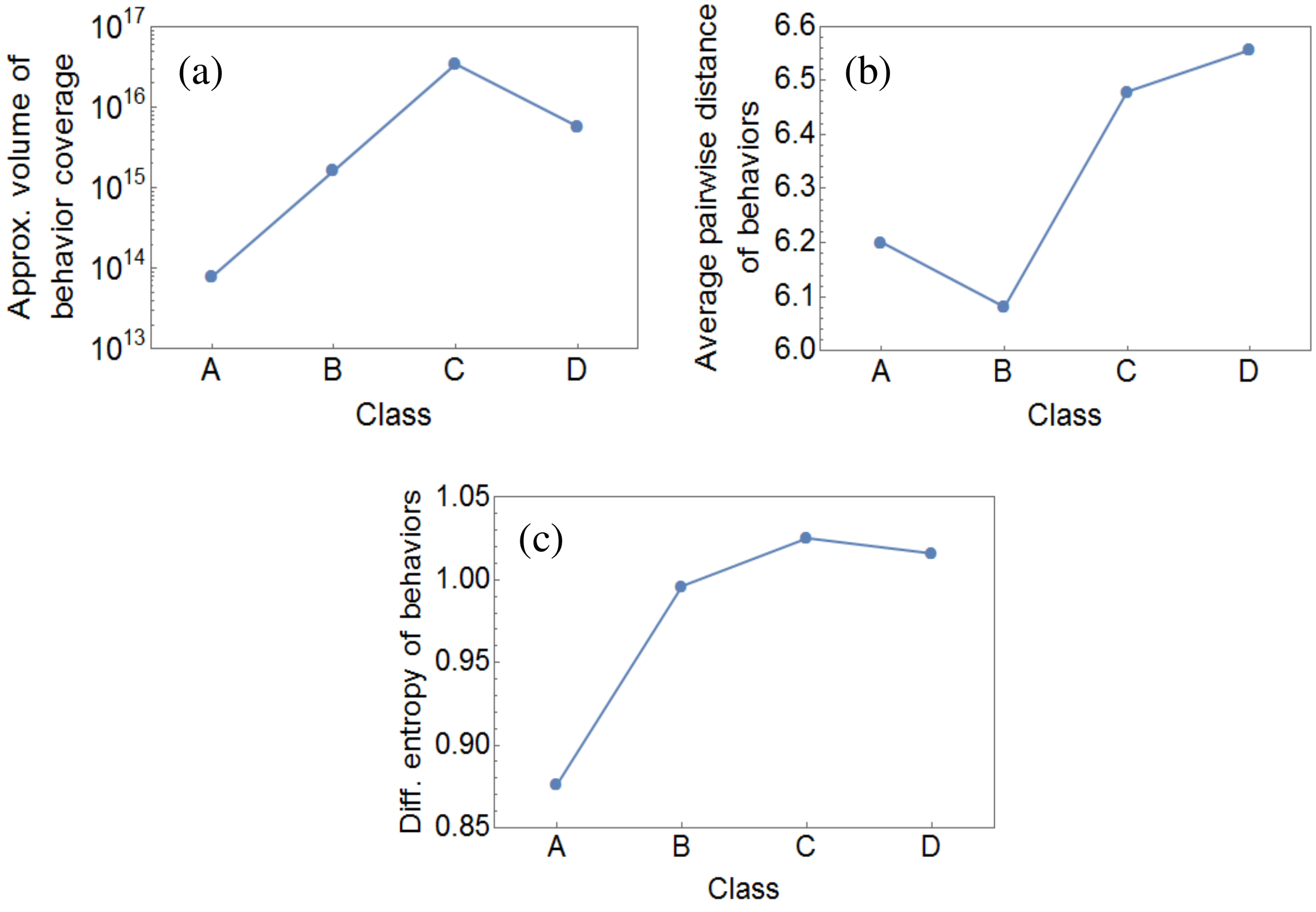}
\caption{Behavioral diversities of morphogenetic collective systems
  measured using three metrics: (a) approximated volume of behavioral
  coverage, (b) average pairwise distance of behaviors, and (c)
  differential entropy of behaviors. In all of the three plots,
  Classes C and D showed greater behavioral diversity than Classes A
  and B.}
\label{fig:diversity}
\end{figure}

As described above, straightforward statistical analysis placed the
properties of Class C/D systems somewhere in between Class A and Class
B, while it did not clarify whether Class C/D systems had any truly
unique properties different from Classes A or B. Therefore, we
conducted more in-depth, meta-level comparative analysis of {\em
  behavioral diversities} between those four classes of morphogenetic
collective systems \cite{sayama2015a}. Behavioral diversities were
measured for each class by computing the approximated volume of
behavior space coverage, the average pairwise distance of two randomly
selected behaviors in the behavioral space, and the differential
entropy \cite{cover2012} of the smoothed behavior distribution. More
details can be found in \cite{sayama2015a}. Results indicated that the
dynamic (re-)differentiation of individual components, which was
unique to Class C/D systems, played a crucial role in increasing the
diversity in possible behaviors of collective systems
(Fig.~\ref{fig:diversity}). This new finding revealed that our
previous interpretation that Class C/D systems would behave more
similarly to Class A than to Class B was not quite accurate. Rather,
the difference between Classes A/B and Classes C/D helped make more
diverse collective structures and behaviors accessible, providing for
a larger ``design space'' for morphogenetic collective systems to
explore.

\section{Evolutionary Design of Morphogenetic Collective Systems}

The remaining question we want to address is how to design novel
self-organizing patterns of morphogenetic collective systems. Unlike
conventional engineered systems for which clear design principles and
methodologies exist, complex systems show nontrivial emergent
macroscopic behaviors that are hard to predict and design from
microscopic rules bottom-up \cite{braha2006}. To design such systems,
the evolutionary approach has been demonstrated to be one of the most
effective means \cite{baryam2003,sayama2014b}. Here we adopt two
different evolutionary approaches: one is interactive evolutionary
computation (IEC) \cite{takagi2001,sayama2009b,bush2011,sayama2015b}
and the other is spontaneous evolution within a simulated artificial
ecosystem \cite{conrad1970,sayama2011a,sayama2011b}.

\begin{figure}[t]
\centering
\includegraphics[width=0.8\columnwidth]{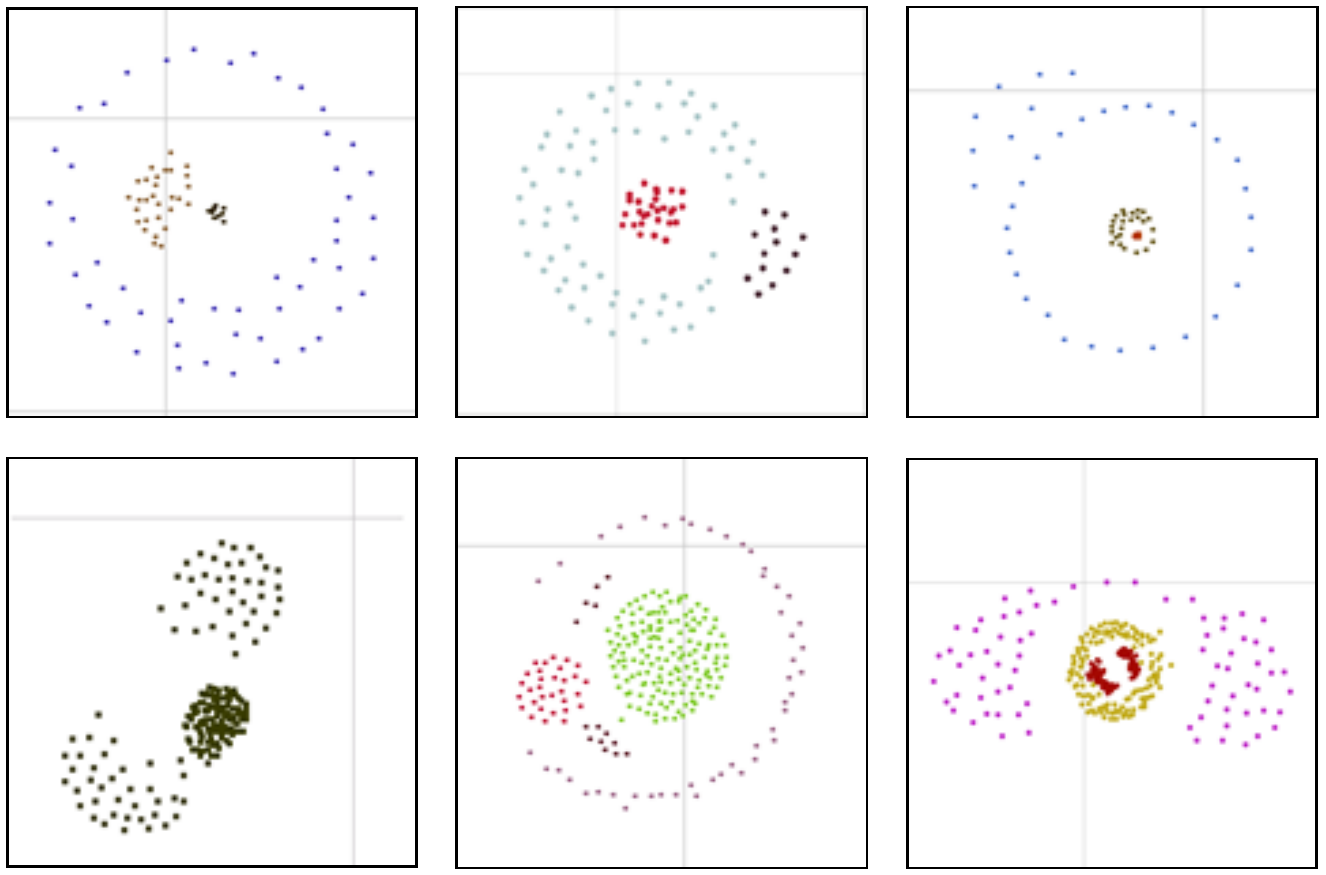}
\caption{Several examples of self-organizing life-like patterns in
  Swarm Chemistry evolved using the interactive evolutionary
  computation approach.}
\label{fig:patterns}
\end{figure}

\begin{figure}[t]
\centering
\includegraphics[width=0.9\columnwidth]{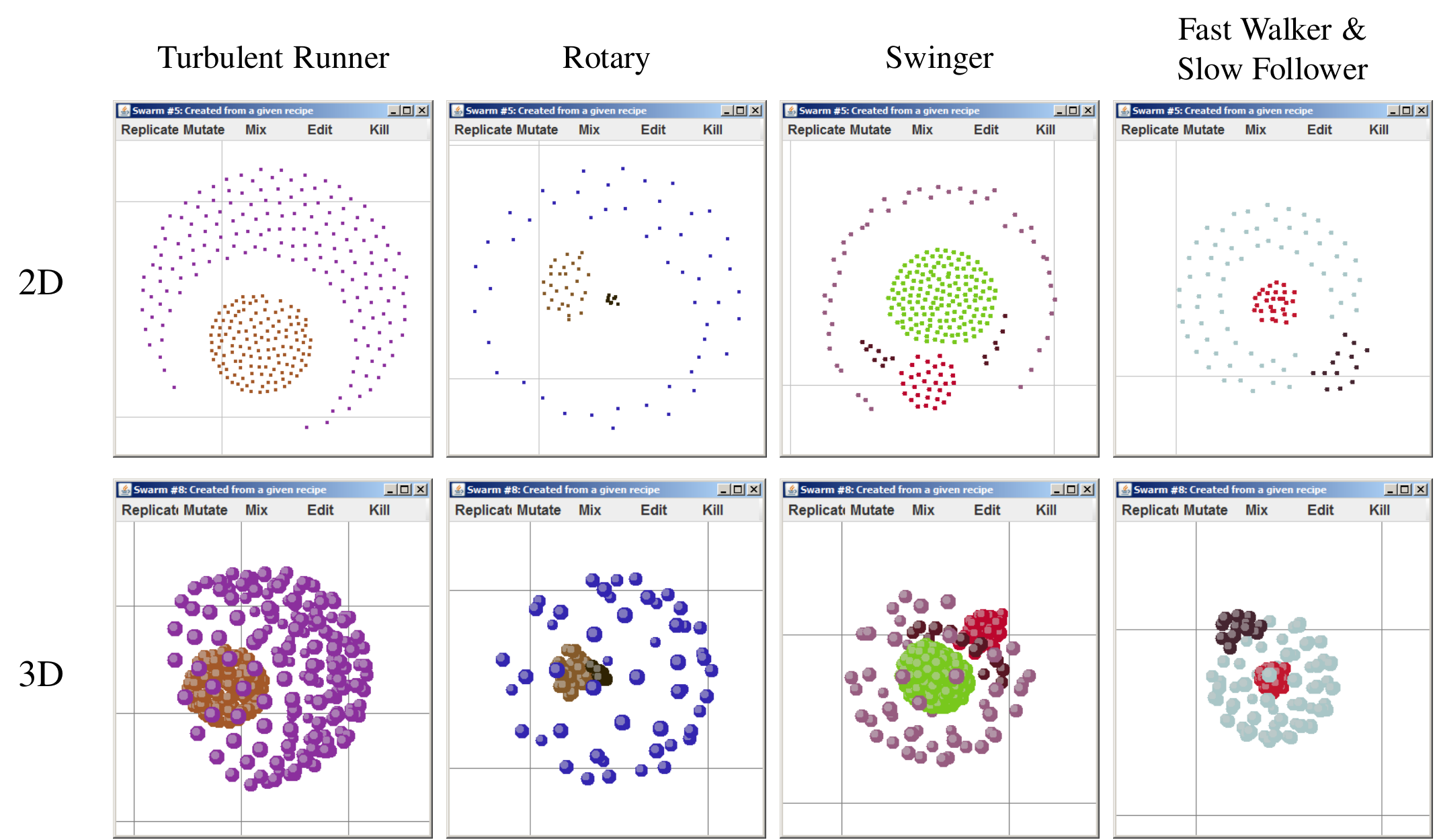}
\caption{Comparison of morphologies between 2D and 3D spaces, both
  developed from identical recipes.}
\label{fig:3D}
\end{figure}

In the IEC approach, we developed a novel IEC framework called
``Hyper-Interactive Evolutionary Computation (HIEC)''
\cite{bush2011,sayama2015b}, in which human users act not only as a
fitness evaluator but also as an active initiator of evolutionary
changes. HIEC was found to be highly effective in exploring the
extremely high dimensional design space of Swarm Chemistry,
discovering a number of nontrivial, life-like morphological patterns
and dynamic behaviors (Fig.~\ref{fig:patterns})\footnote{For more
  evolved patterns, see the Swarm Chemistry website:
  \url{http://bingweb.binghamton.edu/~sayama/SwarmChemistry/}}. We
also found that these designed self-organizing patterns were
remarkably robust against dimensional changes from 2D to 3D
\cite{sayama2012b} (Fig.~\ref{fig:3D}), which is highly unique given
that behaviors of complex systems generally depend heavily on spatial
dimensions in which they develop.

\begin{figure}[t]
\centering
\includegraphics[width=\columnwidth]{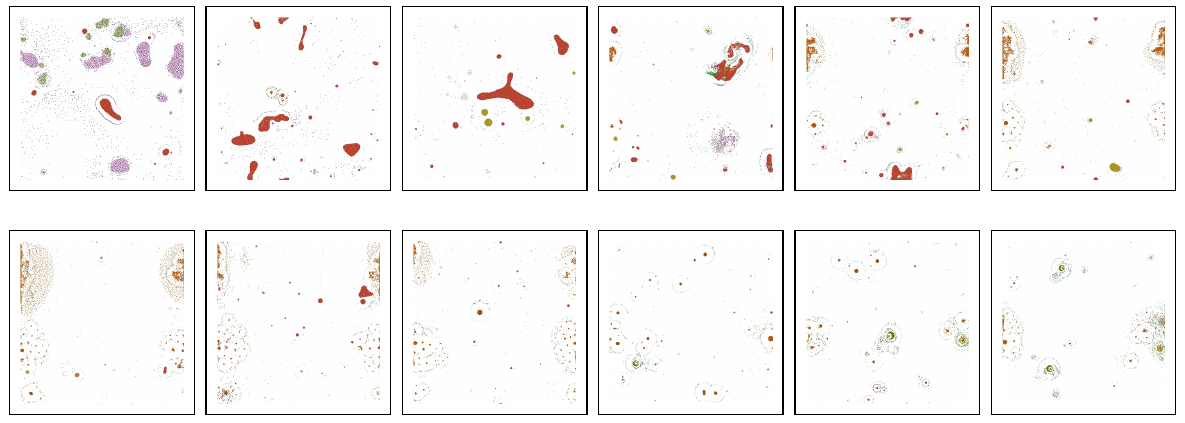}
\caption{Sample simulation run of Evolutionary Swarm Chemistry (from
  \cite{sayama2014b}). Time flows from left to right (the bottom row
  follows the top one).}
\label{fig:evolution}
\end{figure}

Finally, in the spontaneous evolution approach, we replaced the human
users in IEC with microscopic ``physics laws'' that would govern
transmission of recipe information among individual components (as
evolutionary operators acting at local scales) and macroscopic
measurements of ``interestingness'' (as assessments of evolutionary
processes at global scales)
\cite{sayama2011a,sayama2011b}. Specifically, recipe information was
assumed to be transmitted between two colliding particles (with
stochastic mutations possible at a small probability). The direction
of transmission was determined by specific microscopic laws. These
laws were perturbed globally at certain intervals to introduce
variations and thus keep the evolutionary processes active and
ongoing. The interestingness of evolution was measured by spatial
structuredness (i.e., deviation from random homogeneous patterns) and
temporal novelty production rates. More details can be found in
\cite{sayama2011a,sayama2011b}. This spontaneous evolution approach
was shown to be very powerful in continuously producing nontrivial
morphologies. An example is given in Fig.~\ref{fig:evolution}, and
other illustrative evolutionary processes can be found
online\footnote{\url{https://www.youtube.com/user/ComplexSystem/videos}}.

In the meantime, it was also noticed that evolutionary exploration was
much less active in three-dimensional space than in two-dimensional
one \cite{sayama2012c}, despite the robustness of self-organization
against the same dimensional changes. This sensitivity was considered
to be due to the fact that spontaneous evolution heavily relies on
collisions between particles, which would become fundamentally less
frequent in 3D space \cite{polya1921,domb1954}.

\section{Conclusions}

In this chapter, we gave a condensed summary of our recent project
that explored the complexity, development, and evolution of
morphogenetic collective systems. The classification scheme of
morphogenetic collective systems we proposed was among the first that
focuses on functional and interactive capabilities of microscopic
individual components. By orthogonalizing microscopic components'
capabilities with macroscopic system behaviors, one can define a
design space for various forms of morphogenetic collective systems,
which will be useful for both classification of biological collectives
and design of self-organizing artificial collectives.

The numerical simulation results obtained by using Morphogenetic Swarm
Chemistry demonstrated that each of the characteristic properties of
collective systems has unique, distinct effects on the resulting
morphogenetic processes. Heterogeneity of components has quite
significant effects on various properties of the collective systems,
while the ability for individuals to dynamically switch their types
contributes to the spatial coherence, the ability to self-repair, and
the increase of behavioral diversity of those collective systems. Such
behavioral richness would be the necessary ingredient for collective
systems to evolve sophisticated structures and/or functions, which was
partly demonstrated in the evolutionary approaches also discussed
in this chapter.

This short chapter is obviously not sufficient to cover the whole
scope of the project, which also produced several more
application-oriented contributions that were not discussed
here. Interested readers are encouraged to visit our project
website\footnote{\url{http://bingweb.binghamton.edu/~sayama/NSF-RI-MCS/}}.

\section*{Acknowledgments}

This material is based upon work supported by the National Science
Foundation under Grant No.\ 1319152. The author thanks Benjamin James
Bush, Shelley Dionne, Craig Laramee, David Sloan Wilson, and Chun Wong
for their contributions to this project.


\begin{thebibliography}{99}

\bibitem{baryam1997} Bar-Yam, Y.: Dynamics of Complex
  Systems. Addison-Wesley (1997)

\bibitem{benjacob1998} Ben-Jacob, E., Cohen, I., Gutnick, D.L.:
  Cooperative organization of bacterial colonies: from genotype to
  morphotype. Annual Reviews in Microbiology 52(1), 779--806 (1998)

\bibitem{parrish1999} Parrish, J.K., Edelstein-Keshet, L.: Complexity,
  pattern, and evolutionary trade-offs in animal aggregation. Science
  284(5411), 99--101 (1999)

\bibitem{sole2000} Sol\'{e}, R., Goodwin, B.: Signs of Life: How
  Complexity Pervades Biology. Basic Books (2000)

\bibitem{macy2002} Macy, M.W., Willer, R.: From factors to factors:
  computational sociology and agent-based modeling. Annual Review of
  Sociology 28(1), 143--66 (2002)

\bibitem{camazine2003} Camazine, S., et al.: Self-Organization in
  Biological Systems. Princeton University Press (2003)

\bibitem{couzin2003} Couzin, I.D., Krause, J.: Self-organization and
  collective behavior in vertebrates. Advances in the Study of
  Behavior 32, 1--75 (2003)

\bibitem{gershenson2007} Gershenson, C.: Design and control of
  self-organizing systems. CopIt ArXives (2007)

\bibitem{lammer2008} L\"{a}mmer, S., Helbing, D.: Self-control of
  traffic lights and vehicle flows in urban road networks. Journal of
  Statistical Mechanics: Theory and Experiment 2008(04), P04019 (2008)

\bibitem{turner2008} Turner, J.S., Soar, R.C.: Beyond biomimicry: What
  termites can tell us about realizing the living building. First
  International Conference on Industrialized, Intelligent
  Construction, Loughborough University (2008)

\bibitem{turner2011} Turner, J.S.: Termites as models of swarm
  cognition. Swarm Intelligence 5(1), 19--43 (2011)

\bibitem{vicsek2012} Vicsek, T., Zafeiris, A.: Collective
  motion. Physics Reports 517(3), 71--140 (2012)

\bibitem{portugali2012} Portugali, J.: Self-Organization and the
  City. Springer (2012)

\bibitem{doursat2012} Doursat, R., Sayama, H., Michel, O.:
  Morphogenetic engineering: Reconciling self-organization and
  architecture. Morphogenetic Engineering, Springer, pp. 1--24 (2012)

\bibitem{fernandez2014} Fern\'{a}ndez, N., Maldonado, C., Gershenson,
  C.: Information measures of complexity, emergence,
  self-organization, homeostasis, and autopoiesis. Guided
  Self-Organization: Inception, Springer, pp.19--51 (2014)

\bibitem{sayama2015textbook} Sayama, H.: Introduction to the Modeling
  and Analysis of Complex Systems. Open SUNY Textbooks (2015)

\bibitem{sayama2014a} Sayama, H.: Four classes of morphogenetic
  collective systems. Artificial Life 14: Proceedings of the
  Fourteenth International Conference on the Synthesis and Simulation
  of Living Systems, MIT Press, pp. 320--327 (2014)

\bibitem{sayama2009a} Sayama, H.: Swarm chemistry. Artificial Life 15,
  105--114 (2009)

\bibitem{sayama2012a} Sayama, H.: Swarm-based morphogenetic artificial
  life. Morphogenetic Engineering: Toward Programmable Complex
  Systems, Springer, pp.191--208 (2012)

\bibitem{reynolds1987} Reynolds, C.W.: Flocks, herds and schools: A
  distributed behavioral model. ACM SIGGRAPH Computer Graphics 21(4),
  25--34 (1987)

\bibitem{wasserman1994} Wasserman, S., Faust, K.: Social Network
  Analysis: Methods and Applications. Cambridge University Press (1994)

\bibitem{barabasi2016} Barab\'{a}si, A.-L.: Network Science. Cambridge
  University Press (2016)

\bibitem{sayama2010} Sayama, H.: Robust morphogenesis of robotic
  swarms. IEEE Computational Intelligence Magazine 5(3), 43--49 (2010)

\bibitem{sayama2015a} Sayama, H.: Behavioral diversities of
  morphogenetic collective systems. Proceedings of the Thirteenth
  European Conference on Artificial Life (ECAL 2015), MIT Press, p. 41
  (2015)

\bibitem{cover2012} Cover, T.M., Thomas, J.A.: Elements of Information
  Theory. John Wiley \& Sons (2012)

\bibitem{braha2006} Braha, D., Minai, A.A., Bar-Yam Y.: Complex
  Engineered Systems. Springer (2006)

\bibitem{baryam2003} Bar-Yam, Y.: When systems engineering
  fails-toward complex systems engineering. IEEE International
  Conference on Systems, Man and Cybernetics 2003, IEEE,
  pp. 2021--2028 (2003)

\bibitem{sayama2014b} Sayama, H.: Guiding designs of self-organizing
  swarms: Interactive and automated approaches.  Guided
  Self-Organization: Inception, Springer, pp.365--387 (2014)

\bibitem{takagi2001} Takagi, H.: Interactive evolutionary computation:
  Fusion of the capabilities of EC optimization and human
  evaluation. Proceedings of the IEEE 89(9), 1275-1296 (2001)

\bibitem{sayama2009b} Sayama, H., Dionne, S., Laramee, C., Wilson,
  D. S.: Enhancing the architecture of interactive evolutionary design
  for exploring heterogeneous particle swarm dynamics: An in-class
  experiment. Proceedings of the Second IEEE Symposium on Artificial
  Life (IEEE ALIFE 2009), IEEE, pp.85--91 (2009)

\bibitem{bush2011} Bush, B. J., Sayama, H.: Hyperinteractive
  evolutionary computation. IEEE Transactions on Evolutionary
  Computation, 15, 424--433 (2011)

\bibitem{sayama2015b} Sayama, H., Dionne, S. D.: Studying collective
  human decision making and creativity with evolutionary
  computation. Artificial Life, 21, 379--393 (2015)

\bibitem{conrad1970} Conrad, M., Pattee, H.H.: Evolution experiments
  with an artificial ecosystem. Journal of Theoretical Biology 28(3),
  393--409 (1970)

\bibitem{sayama2011a} Sayama, H.: Seeking open-ended evolution in
  Swarm Chemistry. Proceedings of the Third IEEE Symposium on
  Artificial Life (IEEE ALIFE 2011), IEEE, pp.186--193 (2011)

\bibitem{sayama2011b} Sayama, H., Wong, C.: Quantifying evolutionary
  dynamics of Swarm Chemistry. Advances in Artificial Life, ECAL 2011:
  Proceedings of the Eleventh European Conference on Artificial Life,
  MIT Press, pp.729--730 (2011)

\bibitem{sayama2012b} Sayama, H.: Morphologies of self-organizing
  swarms in 3D Swarm Chemistry. Proceedings of the 2012 Genetic and
  Evolutionary Computation Conference (GECCO 2012), pp.577--584 (2012)

\bibitem{sayama2012c} Sayama, H.: Evolutionary Swarm Chemistry in
  three-dimensions. Artificial Life 13: Proceedings of the Thirteenth
  International Conference on the Simulation and Synthesis of Living
  Systems, MIT Press, pp.576--577 (2012)

\bibitem{polya1921} P\'{o}lya, G.: \"{U}ber eine Aufgabe der
  Wahrscheinlichkeitsrechnung betreffend die Irrfahrt im
  Stra\ss{}ennetz. Mathematische Annalen 84, 149--60 (1921)

\bibitem{domb1954} Domb, C.: On multiple returns in the random-walk
  problem. Mathematical Proceedings of the Cambridge Philosophical
  Society 50, 586--591 (1954)

\end{thebibliography}
\end{document}